\documentclass[5p,sort&compress]{elsarticle}

\usepackage{bm}
\usepackage{amsmath}
\usepackage{amssymb}
\usepackage{lineno}
\usepackage[colorlinks=true, allcolors=blue]{hyperref}
\modulolinenumbers[3]

\journal{Journal of Magnetism and Magnetic Materials}

\begin{document}
	
	\begin{frontmatter}
		
		\title{Force acting on a cluster of magnetic nanoparticles in a gradient field: a Langevin dynamics study}
		
		\author[icmm]{Andrey A. Kuznetsov\corref{cor}}
		\ead{kuznetsov.a@icmm.ru}
		\cortext[cor]{Corresponding author}
		
		\address[icmm]{Institute of Continuous Media Mechanics UB RAS, 
			Perm Federal Research Center UB RAS,
			614013, Perm, Russia}
		
		\begin{abstract}
			Magnetophoretic force acting on a rigid spherical cluster of single-domain nanoparticles
			in a constant-gradient weak magnetic field is investigated numerically
			using the Langevin dynamics simulation method. 
			Nanoparticles are randomly and uniformly distributed within the cluster volume.
			They interact with each other via long-range dipole-dipole interactions.
			Simulations reveal that if the total amount of particles in the cluster
			is kept constant, the force decreases with increasing nanoparticle concentration due to
			the demagnetizing field arising inside the cluster.
			Numerically obtained force values with great accuracy can be described 
			by the modified mean-field theory, which was previously successfully used 
			for the description of various dipolar media.
			Within this theory, a new expression is derived, 
			which relates the magnetophoretic mobility of the cluster with
			the concentration of nanoparticles and their dipolar coupling parameter.
			The expression shows that if the number of particles in the cluster is fixed,
			the mobility is a nonmonotonic function of the concentration.
			The optimal concentration values that maximize the mobility 
			for a given amount of magnetic phase and a given dipolar coupling parameter 
			are determined.
		\end{abstract}
		

	\end{frontmatter}

\section{Introduction}
	
	Magnetic beads (or microspheres) are composite objects consisting of magnetic nanoparticles
	embedded in a spherical polymer matrix~\cite{gervald2010synthesis,philippova2011magnetic}. 
	Nanoparticles can be homogeneously distributed within the bead volume,
	placed on its surface or concentrated in its center. 
	Typical sizes of beads are $0.1$--$10~\mu$m.
	The most promising applications of beads are in biotechnology and medicine.
	Among them are magnetic cell separation~\cite{zborowski2007magnetic},
	targeted drug delivery~\cite{dutz2012microfluidic}
	and single-molecule magnetic tweezers~\cite{van2015biological}. 
	
	The physical basis for many applications of magnetic beads 
	is the phenomenon of magnetophoresis, i.e. the motion of 
	magnetic particles under the action of nonuniform magnetic field.
	It is known that the sensitivity of beads to the applied gradient field 
	is among main factors determining their suitability for biomedical purposes~\cite{zborowski2007magnetic}.
	As a result, there are many experimental studies on detailed magnetic characterization
	of different beads from different manufacturers~\cite{xu2012simultaneous,zhou2016magnetic,grob2018magnetic}.
	The present work, on the contrary, uses a simplistic model of the magnetic bead
	to conduct a numerical and analytical study, which will hopefully
	provide new qualitative insights into how the magnetophoretic motion of the bead
	is affected by its size and magnetic content.
	
	\section{Model and methods}
	
	\subsection{Problem formulation}
	
\begin{figure}
		\includegraphics{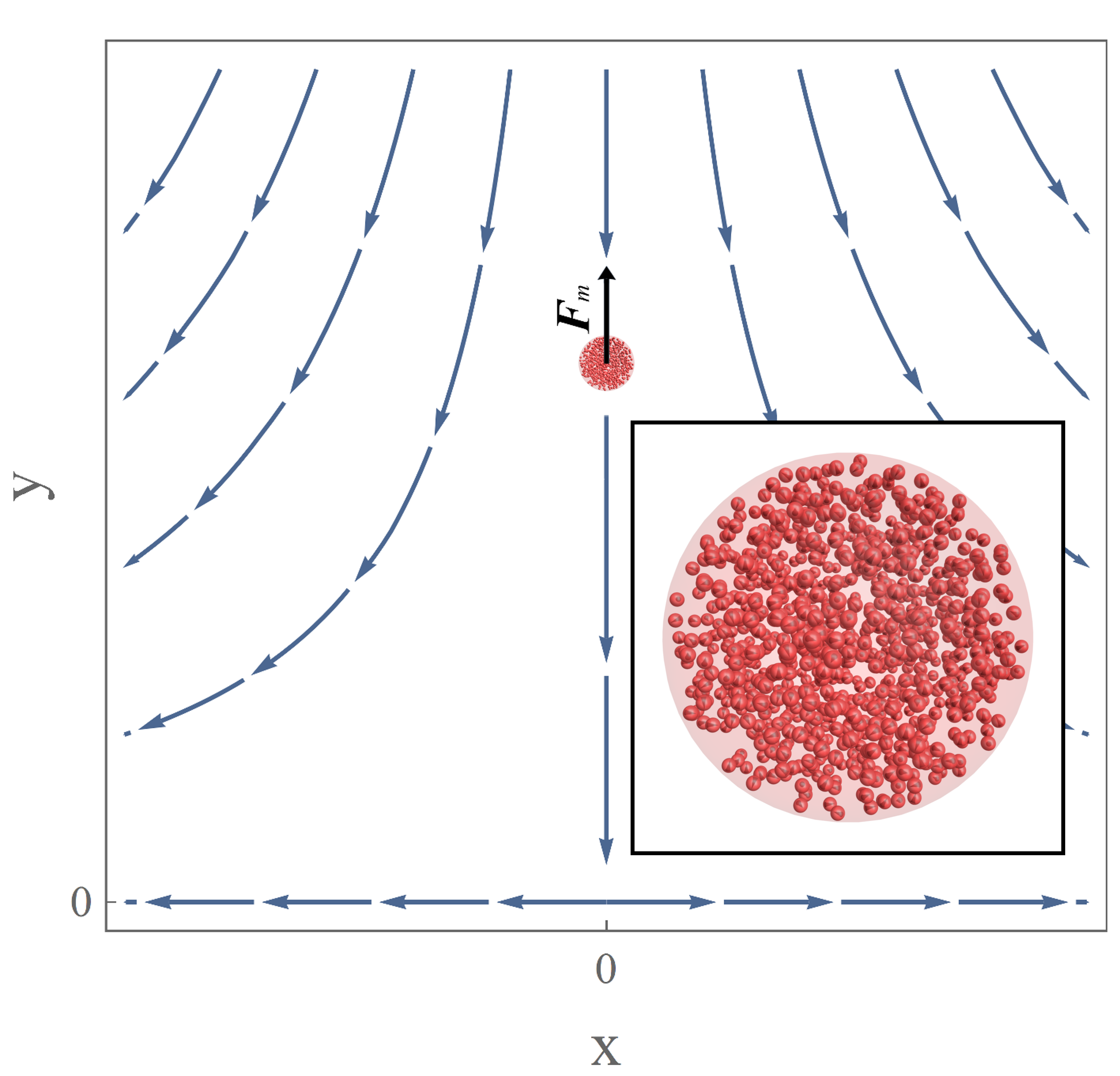}
		\caption{\label{fig:1}
			Schematic sketch of the problem: field lines of the 
			applied quadrupole field $\bm{H}$ and the cluster positioned on the $Y$ axis.
			The inset shows a snapshot of the cluster in its initial state when orientations
			of magnetic moments are random.
			The cluster parameters in the inset are $N = 1000$, $\varphi = 0.05$ and $D \simeq 27d$.
		}
	\end{figure}

	The bead is modeled as a cluster of $N$ identical spherical
	magnetic nanoparticles. The diameter $d$ of particles
	is small enough ($\sim 10$~nm) so that they can be considered as single-domain.
	Each particle has a magnetic moment $\bm{m}$,
	which can rotate freely inside the particle body
	and has a constant magnitude $m = v M_s$,
	where $v = (\pi/6)d^3$ is the particle volume,
	$M_s$ is the saturation magnetization of the particle material.
	Particles are embedded in a rigid nonmagnetic spherical matrix 
	of diameter $D$, their positions are fixed.
	The spatial distribution of particles is random and uniform, 
	no overlapping is allowed.
	Dipole-dipole interactions between particles are taken into account.
	The cluster is placed in a nonmagnetic medium and 
	subjected to a constant nonuniform magnetic field with a gradient $G$. 
	For definiteness, an ideal quadrupole field is considered:
	$\bm{H} = (Gx,-Gy,0)$~\cite{zborowski1999continuous}.
	The schematic sketch of the investigated system is shown in Fig.~\ref{fig:1}.
	The primary task of this study is to determine the magnetic 
	force $\bm{F}_m$ acting on the cluster due to the field for a given~$\bm{R}_c$,
	where $\bm{R}_c = \left(X_c, Y_c, Z_c\right)$ is the location of the cluster center.
	
	Let us introduce a set of appropriate dimensionless parameters
	that determine the cluster behavior.
	The field magnitude can be characterized by the so-called Langevin
	parameter 
	\begin{equation}
	\xi = \frac{\mu_0 m H} {k_B T},
	\end{equation}
	where $\mu_0$ is the vacuum permeability, 
	$k_B$ is the Boltzmann constant, 
	$T$ is the system temperature.
	The Langevin parameter is the ratio between the magnetic (Zeeman) energy of a particle
	in the cluster and the thermal energy $k_B T$.
	The corresponding field vector is $\bm{\xi} =  g(x/d, -y/d, 0)$,
	where $g$  is the dimensionless gradient parameter:
	\begin{equation}
	g = \frac{\mu_0 m G d} {k_B T}.
	\end{equation}
	The intensity of intracluster dipole-dipole 
	interactions can be characterized using the 
	dipolar coupling parameter:
	\begin{equation}
	\lambda = \frac{\mu_0}{4\pi}\frac{m^2}{d^3 k_B T}.
	\end{equation}
	It is known that the ground state of a pair of interacting
	magnetic particles is the ``head-to-tail'' configuration,
	when particles are in close contact and their magnetic moments are collinear~\cite{jacobs1955approach}.
	The dipolar coupling parameter is the ratio between the interaction energy per particle
	in this state and the thermal energy of the system.
	Magnetite nanoparticles, which are typical in biomedical applications,
	can be used as an example to estimate these parameters.
	The saturation magnetization of bulk magnetite is $M_s \simeq 450~\text{kA}\cdot\text{m}^{-1}$ according to Ref.~\cite{rosensweig2002heating},
	but the value $M_s \simeq 350~\text{kA}\cdot\text{m}^{-1}$ is sometimes used for single-domain particles~\cite{schaller2009effective,ilg2017equilibrium}.
	Thus, for magnetite nanoparticles with $d = 10$~nm, the dipolar coupling parameter
	at temperature $T = 300$~K is $\lambda \simeq 1$;
	for particles with $15$~nm, it is $\lambda \simeq 3$--$4.5$.
	Typical gradient values used in the so-called low gradient 
	magnetic separation are $\mu_0 G \sim 10^2~\text{T}\cdot\text{m}^{-1}$~\cite{leong2016working}.
	For magnetite nanoparticles, this corresponds to $g \sim 10^{-4}$.
	Sometimes much larger gradients of the order of $\mu_0 G \sim 10^3~\text{T}\cdot\text{m}^{-1}$ ($g \sim 10^{-3}$)
	are used~\cite{oberteuffer1973high}.
	As for the field magnitude itself, we will here mainly restrict ourself to values $\xi \lesssim 1$.
	In this range, the magnetic response of a nanoparticle ensemble remains linear.
	For 10~nm magnetite particles, $\xi = 1$ corresponds to $H \simeq 15~\text{kA}\cdot\text{m}^{-1}$ (or to $B \simeq 20$~mT).
	This weak field range is relevant for many biomedical diagnostic systems~\cite{xia2006combined,robert2011cell,chen2015one}.
	Besides, this restriction on the field magnitude simplifies 
	the simulation procedure, since now it is possible to neglect the magnetic anisotropy of real single-domain particles.
	It is known that in the general case magnetic anisotropy can significantly 
	effect the magnetization of uniaxial nanoparticles distributed in a solid matrix~\cite{raikher1983magnetization}.
	However, for noninteracting particles with the random easy-axis distribution, 
	the initial slope of the magnetization curve does not depend on the anisotropy energy,
	it is always exactly the same as in the case of isotropic particles~\cite{bean1959superparamagnetism,chantrell1985low}.
	As for interacting uniaxial particles, 
	our recent simulation study~\cite{kuznetsov2018equilibrium}
	also did not find any significant dependency between the weak-field magnetization 
	and the anisotropy energy.
	The last important dimensionless parameter is the volume fraction 
	(volume concentration) of nanoparticles inside the cluster:
	\begin{equation}
	\varphi = \frac{N v } {V} = \frac{N d^3}{D^{3}},
	\end{equation}
	where $V = (\pi/6)D^{3}$ is the cluster volume.
	The notation $x^* = x/d$ will be used for the reduced distance.
	
	\subsection{Langevin dynamics simulations}
	
	In order to accurately take into account 
	the combined effect of the applied field, intracluster interactions and thermal fluctuations 
	on the cluster behavior, the Langevin dynamics method is used.
	The Langevin equation that describes the magnetodynamics of a single-domain particle
	is the stochastic Landau-Lifshitz-Gilbert equation~\cite{brown1963thermal,garcia1998langevin}.
	For the $i$th particle of the simulated cluster it reads
	\begin{equation}\label{eq:llg1}
	\frac{d \bm{m}_i}{d t} = - \gamma \left[\bm{m}_i \times \bm{H}^{tot}_i\right] - \frac{\gamma \alpha}{m} \left[\bm{m}_i \times \left[\bm{m}_i \times \bm{H}^{tot}_i \right]\right],
	\end{equation}
	where $\gamma = \gamma_0 / (1 + \alpha ^2)$, $\gamma_0$ is the gyromagnetic ratio (measured in $\text{m}\cdot\text{A}^{-1}\cdot\text{s}^{-1}$), 
	$\alpha$ is the phenomenological dimensionless damping constant, 
	$\bm{H}^{tot}_i = \bm{H}^{det}_i + \bm{H}^{fl}_i$,
	$\bm{H}^{det}_i$ is the total deterministic field acting on the particle, 
	it is the sum of the applied field and dipolar fields due to all other particles,
	$\bm{H}^{fl}_i$ is the fluctuating thermal field.
	$\bm{H}^{fl}_i(t)$ is a Gaussian stochastic process 
	with the following statistical properties:
	\begin{gather}
	\left\langle H^{fl}_{i,k}(t)\right\rangle =  0, \\
	\left\langle H^{fl}_{i,k}(t_1) H^{fl}_{j,l}(t_2) \right\rangle =  2 {\cal D} \delta_{ij}\delta_{kl}\delta(t_1 - t_2), \\
	{\cal D} = \frac{\alpha k_B T}{\mu_0 m\gamma (1+\alpha^2)},
	\end{gather}
	where $k$ and $l$ are Cartesian indices,
	angle brackets denote a mean value,
	$\delta_{ij}$ is the Kronecker delta,
	$\delta(t)$ is the Dirac delta function,
	${\cal D}$ is the strength of the thermal fluctuations.
	Eq.~(\ref{eq:llg1}) can be rewritten in the dimensionless form:
	\begin{gather}\label{eq:llg}
	\frac{d \bm{e}_i}{d t^*} = - \frac{1}{2 \alpha}\left[\bm{e}_i \times \bm{\xi}^{tot}_i\right] - \frac{1}{2} \left[\bm{e}_i \times \left[\bm{e}_i \times \bm{\xi}^{tot}_i\right]\right],
	\end{gather}
	where the $\bm{e}_i = \bm{m}_i/m$, 
	$t^* = t/\tau_{\cal D}$ is the reduced time,
	$\tau_{\cal D} = \mu_0 m / 2 \alpha \gamma k_B T$ is the characteristic time scale
	of the rotary diffusion of the magnetic moment (typically, $\tau_{\cal D} \sim 10^{-10}$~s~\cite{ilg2017equilibrium}),
	$\bm{\xi}^{tot}_i = \mu_0 m \bm{H}^{tot}_i/k_B T = \bm{\xi}^{det}_i + \bm{\xi}^{fl}_i$,
	\begin{gather}
	\bm{\xi}^{det}_i  = \bm{\xi}_i + 
	\lambda \sum_{j \neq i}^{N} \left[ \frac{3  \bm{r}^*_{ij} (\bm{e}_j \cdot \bm{r}^*_{ij})}{r^{*5}_{ij}} - \frac{\bm{e}_j }{r^{*3}_{ij}} \right], \\
	\left\langle \xi^{fl}_{i,k}(t^*)\right\rangle =  0, \\
	\left\langle \xi^{fl}_{i,k}(t^*_1)\xi^{fl}_{j,l}(t^*_2) \right\rangle =  \frac{4 \alpha^2}{1 + \alpha^2}\delta_{ij}\delta_{kl}\delta(t^*_1 - t^*_2), \label{eq:llhl}
	\end{gather}
	where $\bm{\xi}_i = (gx^*_i, -gy^*_i,0)$,
	$\bm{r}^*_{ij} = \bm{r}^*_{i} - \bm{r}^*_{j}$
	is the vector between centers of particles $i$ and $j$,
	$\bm{r}^*_{i} = (x^*_i, y^*_i, z^*_i)$. 
	
	The input parameters of the simulation are $N$, $\varphi$, $\lambda$, $g$ and 
	$\xi_c = \mu_0 m H_c/k_B T = g\sqrt{X^{*2}_c + Y^{*2}_c}$,
	where $H_c$ is the value of the external field in the center of the cluster.
	In simulations, the cluster is always positioned on the $Y$ axis: $X^*_c = Z^*_c = 0$, $Y^*_c > 0$.
	Using $\xi_c$, the cluster position is determined as $Y^*_c = \xi_c/g$.
	Using $N$ and $\varphi$, the cluster diameter is determined as $D^* = \sqrt[3]{N/\varphi}$. 
	Then the cluster is generated as follows.
	The $i$th particle ($1 \le i \le N$) is randomly placed inside a cube with a side length of $D$
	and with the center located at $(0,Y_c,0)$.
	If after this the particle is outside of the sphere of radius $D/2$ or if 
	it overlaps with previously placed particles (i.e., with particles $j < i$), 
	the position is rejected and the new position is generated. 
	This is repeated until a suitable position is found.
	Then the initial state of $\bm{e}_i$ is chosen at random. 
	Then the state of the particle $i + 1$ is generated according to the same rules.
	After the cluster is generated, the standard Heun scheme~\cite{garcia1998langevin} is used
	for numerical integration of Eqs.~(\ref{eq:llg}-\ref{eq:llhl}). 	
	The damping constant in simulations
	is $\alpha = 0.1$ and the integration time step is $\Delta t^* = 0.002$.
	After every time step, fields $\bm{\xi}^{tot}_i$ are recalculated
	using the current orientations of the particles. 
	Dipolar interaction fields between every pair of 
	particles in the cluster are calculated directly,
	without any truncations or approximations.
	Periodic boundary conditions are not used.
	Position and orientation of the cluster as a whole remain fixed during simulations.
	The following values of input parameters are investigated numerically:
	$0.25 \le \xi \le 2$, $1 \le \lambda \le 7$, $0.05 \le \varphi \le 0.45$, $g = 10^{-3}$,
	$N = 10^2$--$10^3$.
	
	The instantaneous force on a point-like magnetic moment due to external field
	is $\mu_0 (\bm{m} \cdot \nabla)\bm{H}$~\cite{landaulifshitz_edcm}.
	Then, for a quadrupole field, the force on the $i$th particle is
	\begin{equation}
	\bm{F}_{m,i}  = \mu_0 m G(  e_{i,x}, - e_{i,y}, 0).
	\end{equation}
	If the field is large enough, the particle magnetic moment 
	is always aligned with its direction and the magnitude of the force
	has its maximum value $\mu_0 m G$. 
	For modeled systems the condition $gD^* \ll 1$ is typically fulfilled.
	It means that the field magnitude and direction do not significantly
	change within the cluster.
	In this case, all particles in the large field are also collinear with each other 
	and the cluster is saturated, its total magnetic moment is $\mathfrak{M}_{sat} = m N$
	and the force is $F_{sat} = \mu_0 \mathfrak{M}_{sat} G$. 
	A normalized magnetic force then can be introduced as $\bm{f}_m = \bm{F}_m / F_{sat}$.
	For an arbitrary field magnitude, the net external force is calculated in simulations as
	\begin{equation}
	\bm{f}_{m}  = \frac{\left\langle \sum^N_{i = 1} \bm{F}_{m,i}  \right\rangle}{F_{sat}} 
	= \frac{1}{N}\left(\left\langle \sum_{i = 1}^{N} e_{i,x} \right\rangle, - \left\langle \sum_{i = 1}^{N} e_{i,y} \right\rangle, 0\right).
	\end{equation}
	To find this average quantity, the sampling of instantaneous force values starts after 
	the time period of $t = 200 \tau_{\cal D} \sim 10^{-8}$--$10^{-7}$~s.
	In all considered cases, this time is enough for 
	the simulated system to reach equilibrium.
	The total simulated period for each specific set of input parameters is about $2000 \tau_{\cal D}$.
	Note that in biomedical applications, such as magnetic cell separation,
	typical velocities of magnetic microparticles are $10$--$10^2~\mu$m/s~\cite{zborowski2007magnetic}.
	So, the time it takes a microparticle to travel a distance equal to its diameter ($10^{-3}$--$10^{-1}$~s)
	is several orders of magnitude larger than the time required to achieve an equilibrium force value.
	This justifies the neglect of the cluster translational motion in simulations.
	The same reasoning remains valid even if magnetic anisotropy of particles is taken into account.
	It is known that anisotropy slows down the relaxation time of magnetic moments,
	but for 10~nm iron oxide particles this time is still comparable with~$\tau_{\cal D}$~\cite{rosensweig2002heating}.
	The situation can be more complicated for other magnetic materials.
	For example, 10~nm~cobalt ferrite particles have the relaxation time $\gg 1$~s~\cite{claesson2007measurement},
	so the cluster with such particles will presumably remain in a nonequilibrium state during
	its magnetophoretic motion.
	This situation is beyond the scope of the present work.
	For every particular set of input parameters,
	the force values are averaged not only over simulation time but also over ten independent 
	realizations of the cluster.
	These realizations differ in positions of particles and initial orientations of magnetic moments.
	Such averaging can be also considered as an implicit account
	of the cluster rotation which may arise due to small inhomogeneities
	in the particle spatial distribution.
	In practice, force values for different realizations are very close.
	Error bars presented on the plots below show 95\% confidence intervals for calculated averages.
	
	\subsection{Analytical solution}
			
	A much more common approach to the problem at hand 
	is to consider the cluster as a homogeneous paramagnetic sphere of volume $V$.
	In this approximation, 
	if the gradient is relatively small ($gD^* \ll 1$) and 
	if the cluster and the surrounding medium are linearly magnetizable,
	the magnetic force on the cluster due to external field $\bm{H}$ is~\cite{cummings1976capture,rinaldi2009invariant}:   
	\begin{equation} 
	\bm{F}_{m} = \frac{3}{2}\mu_s \frac{\mu_c - \mu_s}{\mu_c + 2\mu_s} V \nabla H^2, \label{eq:theor_force_general_0}
	\end{equation}
	where $\mu_c$ and $\mu_s$ are absolute magnetic permeabilities of the cluster material 
	and the surrounding medium, correspondingly.
	For the nonmagnetic medium, $\mu_s = \mu_0$ and Eq.~(\ref{eq:theor_force_general_0}) reduces to
	\begin{gather} 
	\bm{F}_{m} = \mu_0 \chi V \nabla \left(H^2/2\right), \label{eq:theor_force_general} \\
	\chi = \frac{\chi_c}{1 + \chi_c/3}, 
	\end{gather}
	where $\chi_c = \mu_c/\mu_0 - 1$ is the initial susceptibility of the cluster material.
	To elaborate on the meaning of the quantity $\chi$,
	let us first consider an elongated cylindrical sample homogeneously filled with the magnetic material
	with the susceptibility $\chi_c$.
	If a weak uniform magnetic field $H$ is applied along the main axis
	of the cylinder, then the relation between the sample magnetization and the field
	is $M = \chi_c H$.
	For a spherical sample, the situation becomes more complicated.
	Now $\chi_c$ describes the relation between the magnetization
	and the magnetic field \textit{inside} the sample $H_{int}$,
	i.e. $M = \chi_c H_{int}$.
	The internal field does not coincide with the applied one,
	the difference between two fields is called the demagnetizing field.
	It is created by the surface divergence of the sample's own magnetization~\cite{joseph1965demagnetizing}.
	For a sphere, the relation between applied and internal fields is
	$H = H_{int} + M/3$.  
	It is easy to see, that for a sphere $M = \chi_c H/(1 + \chi_c/3) = \chi H$.
	So, $\chi$ can be considered as the cluster effective susceptibility.
	The total magnetic moment of the cluster is $\mathfrak{M} = \chi V H$.
	
	For a quadrupole field, the force Eq.~(\ref{eq:theor_force_general}) can be rewritten in the normalized form
	\begin{align} \label{eq:theor_force}
	\bm{f}_{m} &= \frac{\mu_0 \chi V G^2 }{F_{sat}} \left(X_c, Y_c, 0\right) \nonumber\\
	&= \frac{\mu_0 \chi V G^2 \xi_c d}{F_{sat} g  }  \left(\frac{X^*_c}{\sqrt{X^{*2}_c + Y^{*2}_c}}, \frac{Y^*_c}{\sqrt{X^{*2}_c + Y^{*2}_c}}, 0\right)\nonumber \\
	&= \frac{ \chi  \xi_c }{3 \chi_L} \left(\frac{X^*_c}{\sqrt{X^{*2}_c + Y^{*2}_c}}, \frac{Y^*_c}{\sqrt{X^{*2}_c + Y^{*2}_c}}, 0\right), 
	\end{align}
	where $\chi_L$ is the so-called 
	Langevin susceptibility, which describes the initial magnetic response 
	of an ideal paramagnetic gas~\cite{bean1959superparamagnetism}:
	\begin{equation}
	\chi_L = \frac{\mu_0 m^2 N} {3 k_B T V} = 8 \lambda \varphi
	\end{equation}
	Now the only unknown quantity is the initial susceptibility $\chi_c$
	of the cluster material.
	In our case, this is a solid dispersion of interacting single-domain nanoparticles.  
	To estimate $\chi_c$, we will use the so-called modified mean-field theory (MMFT).
	This approach was first proposed for the description of static magnetic properties 
	of concentrated ferrofluids~\cite{pshenichnikov1996magneto,ivanov2001magnetic}.
	MMFT also showed its effectiveness in the description of other media containing magnetic nanoparticles,
	such as ferrogels~\cite{wood2011modeling} and  magnetic emulsions~\cite{ivanov2012nonmonotonic}.
	It was shown in Refs.~\cite{kuznetsov2018equilibrium,pshenichnikov2000equilibrium} 
	that MMFT is able to successfully describe the initial susceptibility of 
	nanoparticles embedded in a solid nonmagnetic matrix.
	According to MMFT, the susceptibility of an ensemble of single-domain nanoparticles
	is given by
	\begin{equation}\label{eq:susc0}
	\chi_c = \chi_L(1 + \chi_L/3).
	\end{equation}
	Then the cluster effective susceptibility takes the form 
	\begin{equation}\label{eq:susc}
	\chi = \chi_L\frac{1 + \chi_L/3}{1 + \chi_L/3 + \chi^2_L/9}.
	\end{equation}
	Eqs.~(\ref{eq:theor_force}) and (\ref{eq:susc}) completely determine
	the magnetic force acting on the cluster. 
	Their applicability range is to be tested via numerical simulations.
		
	\section{Results and discussion}
	
	\subsection{Magnetic force}
	
		\begin{figure*}
			\includegraphics{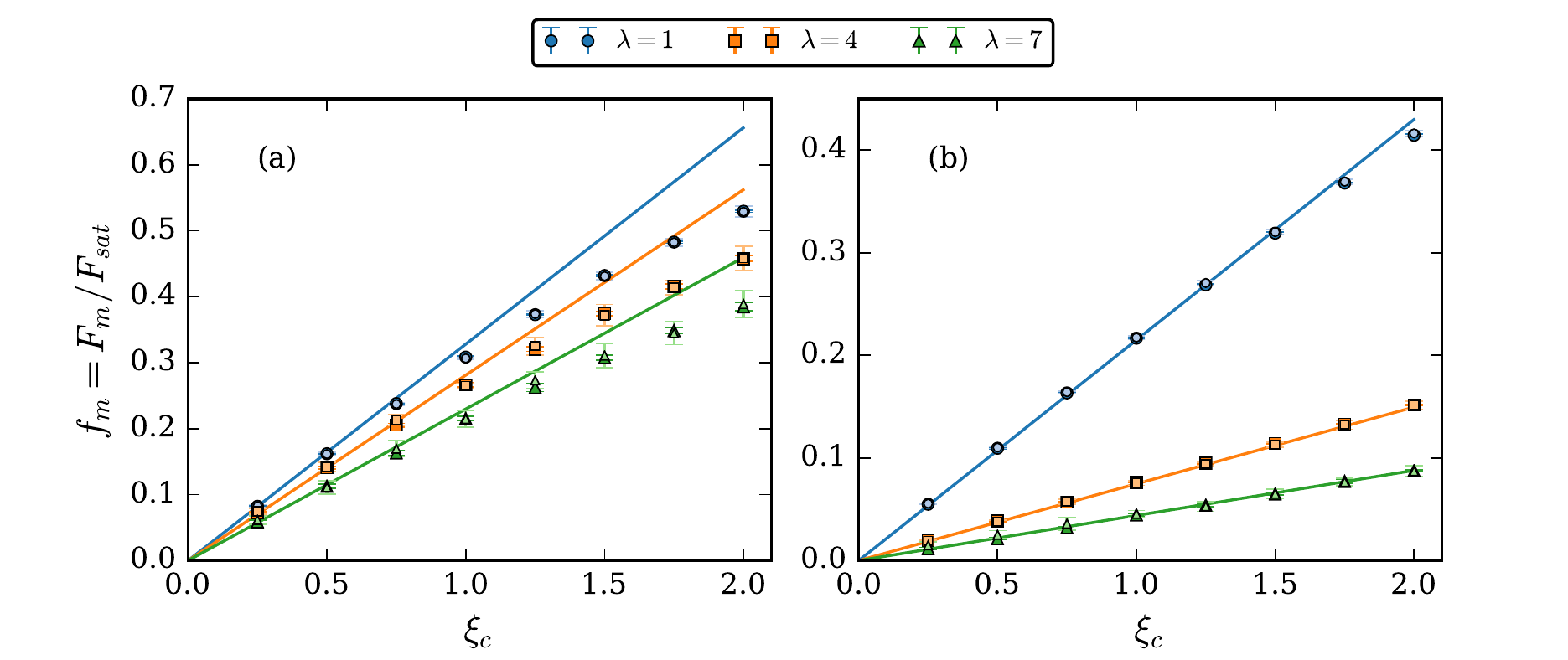}
			\caption{\label{fig:2}
				Normalized magnetic force on the cluster vs. 
				magnetic field in the center of the cluster. 
				$\varphi = 0.05$ (a) and 0.4 (b).
				Symbols are simulations results for $g = 0.001$.
				Different symbols correspond to different dipolar coupling parameters (see legend).
				Larger and darker symbols are for $N = 1000$,
				smaller and lighter symbols are for $N = 100$.
				Solid lines are analytical predictions from Eqs.~(\ref{eq:theor_force}) and (\ref{eq:susc})
				for the same dipolar coupling parameters.
			}
		\end{figure*}
	
	One the simulation results is that the force acting on the cluster positioned on the $Y$ axis
	is directed predominantly along this axis,
	the average $x$-component of the force is zero within the error bar.
	Fig.~\ref{fig:2} illustrates dependencies of the magnetic force magnitude
	on the magnetic field intensity~$\xi_c$.
	It is seen that Eqs.~(\ref{eq:theor_force}) and (\ref{eq:susc}) accurately describe
	simulation results in the weak field limit.
	With increasing field, the growth of the force slows down due to 
	the fact that the cluster magnetization curve is nonlinear -- its magnetic moment
	cannot be larger than $\mathfrak{M}_{sat}$ and the force cannot be larger
	than the corresponding value~$F_{sat}$.
	It is noteworthy that the field range, where the linear response 
	assumption is valid, increases with increasing particle concentration.
	In Fig.~\ref{fig:2}a, which corresponds to $\varphi = 0.05$,
	nonlinearity becomes noticeable already at $\xi \simeq 1$,
	but in Fig.~\ref{fig:2}b ($\varphi = 0.4$) the linear law Eq.~(\ref{eq:theor_force})
	is valid up to $\xi = 2$.
	Fig.~\ref{fig:2} gives simulation results for clusters of different sizes, 
	$N = 100$ and $N = 1000$. 
	Simulation points for two cases are very close and this is an encouraging result.
	Due to limited computational resources, 
	we only investigate clusters
	with $N \sim 10^3$, which at the lowest considered concentrations $\varphi \sim 0.1$ 
	have a diameter of a few tenths of a micron.  
	But the weak dependency of the cluster reduced properties on 
	its size indicates that obtained results should remain relevant for larger 
	structures with $D \sim 1$--10~$\mu$m.
	
	\begin{figure}
		\includegraphics{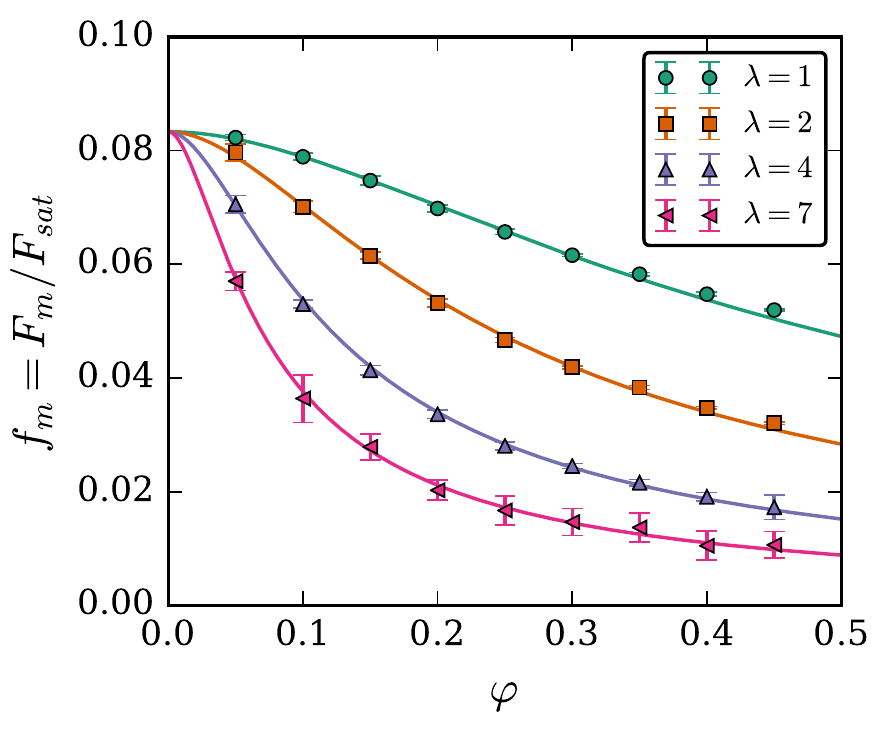}
		\caption{\label{fig:3}
			Normalized magnetic force on the cluster vs. 
			volume fraction of particles in the cluster. 
			$\xi_c = 0.25$.
			Symbols are simulations results for $N = 500$ and $g = 0.001$.
			Different symbols correspond to different dipolar coupling parameters (see legend).
			Curves are analytical predictions from Eq.~(\ref{eq:theor_force}) and (\ref{eq:susc})
			for the same dipolar constants.
		}
	\end{figure}
	Simulations also show that the force acting on an $N$-particle 
	cluster is smaller for more concentrated clusters.
	It is clearly seen in Fig.~\ref{fig:3}.
	In the limit $\varphi \rightarrow 0$, the normalized force is $f_m = \xi_c/3$.
	With increasing $\varphi$, the force starts a nonlinear decline,
	which is more pronounced at larger coupling parameters $\lambda$.
	At $\lambda = 7$ and $\varphi = 0.45$, the force drops by almost an order of magnitude.
	MMFT accurately describes simulation results for all considered values 
	of $\varphi$ and $\lambda$ and can be used to analyze the observed behavior.
	The total magnetic moment of the cluster and the magnetic force Eq.~(\ref{eq:theor_force_general}) are 
	proportional to the quantity $\chi V = (\chi / \varphi) V_m$,
	where $V_m = v N$ is the total amount of magnetic material in the cluster.
	In the case when $V$ is fixed, the force is controlled by
	the susceptibility $\chi$, which is a measure of the magnetic response per unit volume.
	But if $V_m$ is fixed (this is the case in simulations), the force is determined
	by $\chi/\varphi$, which is a measure of the magnetic response per particle.
	If intracluster interactions between particles are neglected (the Langevin approximation),
	the susceptibility is given by the Langevin value $\chi = \chi_L$,
	which grows linearly with the concentration $\varphi$.
	As a result, for a given $V_m$, the quantity $\chi/\varphi$ and hence the force do not depend on the particle concentration.
	The force always equals to the zero-concentration value $F_m = (\xi_c/3) F_{sat} = (8 \lambda V_m) \mu_0 G H_c $.
	Eq.~(\ref{eq:susc0}) goes beyond the Langevin approximation and 
	takes into account the fact that dipole-dipole interactions between 
	an arbitrary particle and its local surroundings,
	on average, help the particle to align with the field. 
	$\chi_c$ grows quadratically with the concentration and $\chi_c/\varphi$ grows linearly. 
	Eq.~(\ref{eq:susc}) additionally takes into account the demagnetizing
	field, which is the long-range effect of dipole-dipole interactions.
	This field, in accordance with its name, weakens the response of an arbitrary particle 
	to the applied field.
	The demagnetizing field is proportional to $\chi$, which grows slower than linearly with $\varphi$
	and bounded from above by the value $\chi = 3$ (see Fig.~\ref{fig:4} below).
	Consequently, at fixed $V_m$ and large $\varphi$, the quantity $\chi/\varphi$ and hence the force decrease 
	hyperbolically with the concentration,  $F_m = (\xi_c/8 \lambda \varphi) F_{sat} = (3V_m/\varphi)\mu_0 G H_c$. 
	
	\subsection{Magnetophoretic mobility}

	\begin{figure}
		\includegraphics{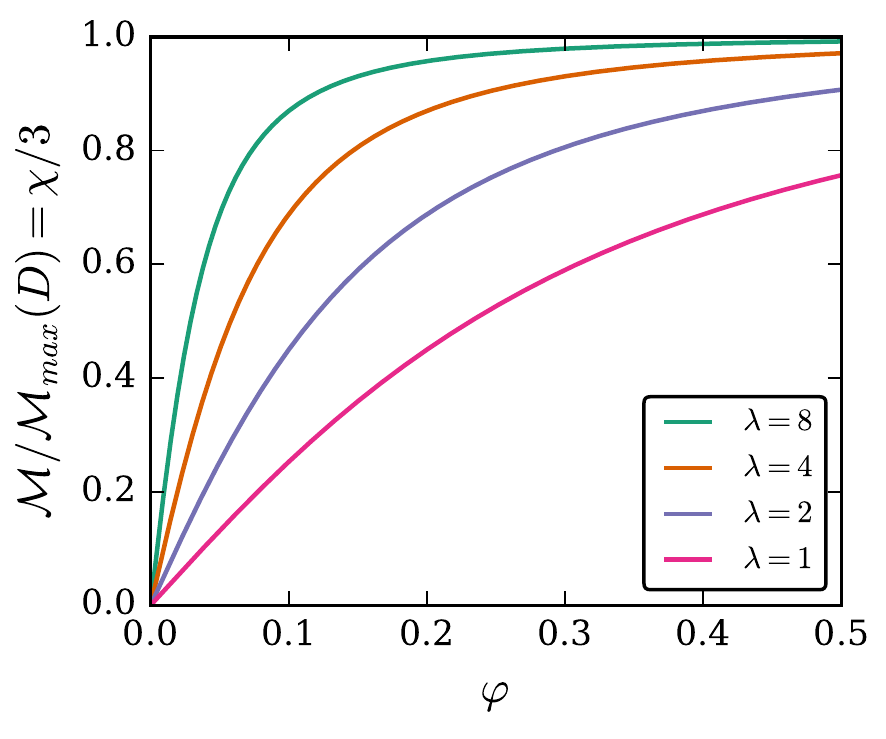}
		\caption{\label{fig:4}
			Normalized magnetophoretic mobility of the cluster vs. 
			volume fraction of particles in it [Eq.~(\ref{eq:mob})]. 
			The cluster diameter is fixed ($D = \text{const}$).
			Different curves correspond to different dipolar coupling parameters,
			from bottom to top: $\lambda = 1$,~2,~4~and~8.
		}
	\end{figure}
	\begin{figure}
		\includegraphics{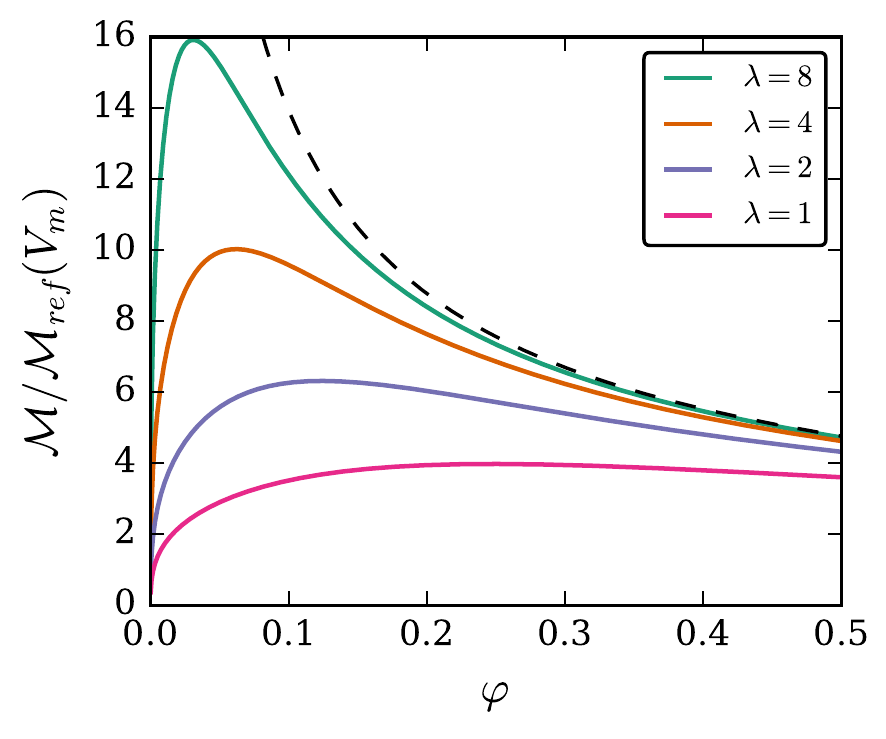}
		\caption{\label{fig:5}
			Magnetophoretic mobility of the cluster vs. 
			volume fraction of particles in it [Eq.~(\ref{eq:mob})]. 
			The total amount of magnetic phase in the cluster is fixed ($V_m = \text{const}$).
			Different solid curves correspond to different dipolar coupling parameters,
			from bottom to top: $\lambda = 1$, 2, 4 and 8.
			Dashed line is the high-concentration asymptote~$3/\varphi^{2/3}$.
		}
	\end{figure}
	
	The analytical model based on MMFT
	shows very good agreement with the simulation results in a wide range of
	interaction parameters $\varphi$ and $\lambda$. 
	Potentially, the model can be used for an accurate description of 
	the cluster magnetophoresis at temporal and spatial scales 
	that are not easily accessible via direct nanoscale simulations. 
	For example, the model can be used to obtain a universal expression
	for the so-called magnetophoretic mobility.
	It is known that magnetic microparticles moving in a viscous nonmagnetic liquid
	with time attain a constant velocity value $u$, which is determined 
	by the balance between the magnetic force $\bm{F}_m$
	and the drag force $\bm{F}_d$~\cite{zborowski2007magnetic}.
	The latter for spherical particles with low Reynolds numbers is given by the Stokes's law:
	\begin{equation}\label{eq:stokes}
	\bm{F}_d = - 3 \pi \eta D \bm{u},
	\end{equation}
	where $\eta$ is the viscosity of the suspending liquid.
	In the general case, the drag force should contain the hydrodynamic 
	diameter of the cluster $D_h$, which can be larger than $D$ 
	if the cluster core is covered by some nonmagnetic shell. 
	But here we assume that two diameters coincide.
	From the balance condition $\bm{F}_m = - \bm{F}_d$ one obtains
	\begin{equation}
	\bm{u} = \frac{ \chi V}{3 \pi \eta D}\bm{S}_m,
	\end{equation}
	where $\bm{S}_m = \mu_0 (\bm{H} \cdot \nabla) \bm{H} = \nabla \left( \mu_0 H^2 / 2\right)$ is known
	as the magnetophoretic driving force~\cite{zborowski2007magnetic,grob2018magnetic}.
	The proportionality factor ${\cal M} = \chi V / 3 \pi \eta D$
	is the cluster magnetophoretic mobility.
	Using Eq.~(\ref{eq:susc}), the mobility can be rewritten as
	\begin{align}\label{eq:mob}
	{\cal M} & = \frac{\chi D^2}{18 \eta} \nonumber\\
	&= \frac{\chi}{\varphi^{2/3}} \frac{1}{18 \eta} \left[\frac{6}{\pi}V_m\right]^{2/3} \nonumber\\
	&= \frac{\chi_L^{1/3}(1+\chi_L/3)}{1 + \chi_L/3 + \chi_L^2/9} \frac{4 \lambda^{2/3}}{18 \eta} \left[\frac{6}{\pi}V_m\right]^{2/3}.
	\end{align} 
	It should be emphasized that Eq.~(\ref{eq:mob}) is not specifically tied to 
	a quadrupole field considered earlier.
	However, it still assumes that the field magnitude is small 
	and the cluster response is linear.
	According to Eq.~(\ref{eq:mob}), the maximum possible mobility 
	for a given diameter $D$ is ${\cal M}_{max}(D) = D^2 / 6 \eta$.
	For $D = 1~\mu$m and $\eta~=~10^{-3}~\text{Pa}\cdot\text{s}$, this value is 
	${\cal M}_{max} = 1.67 \cdot 10^{-10}~\text{m}^3(\text{T}\cdot\text{A}\cdot\text{s})^{-1}$.
	The concentration dependency of the normalized mobility ${\cal M/M}_{max}(D) = \chi/3$ 
	simply repeats the concentration dependency of the susceptibility. 
	Dependencies for different $\lambda$ are shown in Fig.~\ref{fig:4}.
	Since the diameter is fixed, the friction coefficient $3\pi\eta D$ is constant and the increase
	in the concentration only leads to the slow increase in the magnetic force and hence in the mobility.
	A more complex concentration dependency is observed if, instead of $D$, the total amount 
	of magnetic material $V_m$ is fixed.
	The quantity ${\cal M}_{ref}(V_m) = \left(6V_m/\pi\right)^{2/3}/18\eta$ 
	may be chosen as a reference mobility value for this case.
	For $N = 10^5$, $d = 10$~nm and $\eta~=~10^{-3}~\text{Pa}\cdot\text{s}$, this value is
	${\cal M}_{ref} = 1.2 \cdot 10^{-11}~\text{m}^3(\text{T}\cdot\text{A}\cdot\text{s})^{-1}$.
	Concentration dependencies of ${\cal M/M}_{ref}(V_m)$ at different $\lambda$ 
	are given in Fig.~\ref{fig:5}.
	It is seen that dependencies are nonmonotonic, for every $\lambda$ there is 
	an optimal concentration value at which the mobility is maximal.
	The corresponding value of $\chi_L$ can be found by
	solving $(\partial {\cal M} /\partial \chi_L)_{\lambda, V_m} = 0$.
	It gives the optimal value of the Langevin susceptibility 
	$\chi_{L,opt} \simeq 1.9813$. Then the optimal volume fraction
	and diameter are
	\begin{equation}
	\varphi_{opt} = \frac{1.9813}{8 \lambda}, \:\: D_{opt} = \left[\frac{6}{\pi}\frac{V_m}{\varphi_{opt}}\right]^{1/3}.
	\end{equation}
	The mobility under these optimal conditions is ${\cal M}_{opt} \simeq 4 \lambda^{2/3} {\cal M}_{ref}(V_m)$.
	The observed nonmonotonic dependency can be explained as follows.
	In the limit $\varphi \rightarrow 0$,
	the parameter $(\chi/\varphi)V_m$, which controls the magnetic force, 
	has a definite finite value $8\lambda V_m$.
	On the contrary, the cluster diameter and hence the friction coefficient
	are infinitely large, so the mobility in this limit is zero.
	With increasing concentration,
	the friction coefficient decreases as $ \sim 1/\varphi^{1/3}$
	and the mobility initially increases as ${\cal M} = 8\lambda \varphi^{1/3} {\cal M}_{ref}(V_m)$.
	But at $\varphi > \varphi_{opt}$, the magnetic force decrease becomes 
	hyperbolic (due to strong demagnetizing fields) and dominates over
	the drag decrease. As a result, the mobility eventually falls down as 
	${\cal M} =( 3/\varphi^{2/3} ){\cal M}_{ref}(V_m)$.
	
	\section{Conclusions}
	
	In this work, the force acting on a polymer magnetic bead 
	in a constant-gradient field is calculated by means of the Langevin dynamics method. 
	The bead is modeled as a spherical rigid cluster of 
	randomly distributed single-domain particles.
	The magnitude of the applied field is typically small enough
	so that the cluster magnetization remains a linear function of the field.
	It is demonstrated that if the total number of particles
	in the cluster is fixed, the increase in the particle concentration leads
	to the nonlinear decrease in the force magnitude.
	The reason for this is the demagnetizing field inside the cluster,
	which weakens the response of an arbitrary particle to the 
	applied field and hence decreases the cluster net average magnetic moment.
	It is also shown that the cluster can be successfully represented 
	as a single paramagnetic particle whose magnetization
	obeys MMFT. The theory describes numerically obtained force values
	with great accuracy in a broad range of simulation parameters.
	Within MMFT, a new universal formula is obtained for the magnetophoretic mobility
	of an isolated cluster moving in a viscous nonmagnetic liquid. 
	The formula shows that for a given number of particles and 
	a given dipolar coupling parameter there is an optimal concentration
	value (and hence an optimal diameter) for which the mobility is maximal. 
	Below this value, the mobility becomes smaller due to the increase of the cluster friction coefficient;
	above this value, the mobility becomes smaller due to the discussed decrease of the magnetic force. 
	
	In future, we hope to investigate a more general problem 
	when nanoparticles do not occupy the whole bead, but instead distributed
	only in an outer spherical shell surrounding a nonmagnetic polymer core.
	
	\section{Acknowledgments}
	
	The work was supported by Russian Science Foundation (project No. 17-72-10033).
	Calculations were performed using the ``URAN'' supercomputer of IMM UB RAS. 
	
	
	\bibliography{lib}
	
\end{document}